\documentstyle[sprocl]{article}
\input{psfig}
\def\Journal#1#2#3#4{{#1} {\bf #2}, #3 (#4)}
\def\NPB{{\em Nucl. Phys.} B}
\def\PLB{{\em Phys. Lett.}  B}
\def\PRL{\em Phys. Rev. Lett.}
\def\CMP{\em Comm. Math. Phys.}
\def\PRD{{\em Phys. Rev.} D}

\def\ra{\rightarrow}
\def\half{{\scriptstyle{1\over 2}}}
\def\zahlen{{\rm Z \!\! Z}}
\def\sn{{\scriptstyle{1-}}}
\def\Tr{{\rm Tr}}
\def\cO{{\cal O}}
\def\cF{{\cal F}}
\def\cD{{\cal D}}

\def\be{\begin{equation}}
\def\ee{\end{equation}}
\def\bea{\begin{eqnarray}}
\def\eea{\end{eqnarray}}

\newcommand{\basispl}{
   \put(-.5,-.5){\line(1,0){1}}
   \put(.5,-.5){\line(0,1){1}}
   \put(.5,.5){\line(-1,0){1}}
   \put(-.5,.5){\line(0,-1){1}}}
\newcommand{\basisar}{
   \put(0,-.5){\vector(1,0){0}}
   \put(.5,0){\vector(0,1){0}}
   \put(0,.5){\vector(-1,0){0}}
   \put(-.5,0){\vector(0,-1){0}}}
\newcommand{\plaq}{\setlength{\unitlength}{.5cm}\raisebox{-.2cm}{
   \begin{picture}(1.2,1.2)(-.6,-.6)
   \basispl\basisar
   \put(-.5,-.5){\circle*{.2}}
   \put(-.55,-.55){\makebox(0,0)[tr]{\footnotesize $x$}}
   \put(-.55,0){\makebox(0,0)[r]{\footnotesize $\nu$}}
   \put(0,-.55){\makebox(0,0)[t]{\footnotesize $\mu$}}
   \end{picture}}}
\newcommand{\twooneplaq}{\setlength{\unitlength}{.5cm}
   \raisebox{-.2cm}{
   \begin{picture}(2.2,1.2)(-1.1,-.6)
   \put(-1,-.5){\line(1,0){2}}
   \put(-1,.5){\line(1,0){2}}
   \put(-1,-.5){\line(0,1){1}}
   \put(1,-.5){\line(0,1){1}}
   \multiput(-1,-.5)(1,0){3}{\circle*{.2}}
   \multiput(-1,.5)(1,0){3}{\circle*{.2}}
   \end{picture}}}
\newcommand{\plaqa}{\setlength{\unitlength}{.5cm}\raisebox{-.2cm}{
   \begin{picture}(1.2,1.2)(-.6,-.6)
   \basispl
   \put(-.5,-.5){\circle*{.2}}
   \put(-.5,.5){\circle*{.2}}
   \put(.5,-.5){\circle*{.2}}
   \put(.5,.5){\circle*{.2}}
   \end{picture}}}
\newcommand{\hookplaq}{\setlength{\unitlength}{.5cm}
   \raisebox{-.3268cm}{
   \begin{picture}(1.7071,1.7071)(-.7071,-.7071)
   \put(0,0){\line(0,1){1}}
   \put(0,1){\line(1,0){1}}
   \put(1,1){\line(0,-1){1}}
   \put(-.7071,-.7071){\line(1,0){1}}
   \put(0,0){\line(-1,-1){.7071}}
   \put(1,0){\line(-1,-1){.7071}}
   \multiput(0,0)(1,0){2}{\circle*{.2}}
   \multiput(0,1)(1,0){2}{\circle*{.2}}
   \multiput(-.7071,-.7071)(1,0){2}{\circle*{.2}}
   \multiput(0,0)(.25,0){4}{\circle*{.03}}
   \end{picture}}}
\newcommand{\cornplaq}{\setlength{\unitlength}{.5cm}
   \raisebox{-.3268cm}{
   \begin{picture}(1.7071,1.7071)(-.7071,-.7071)
   \put(-.7071,-.7071){\line(0,1){1}}
   \put(0,1){\line(1,0){1}}
   \put(1,1){\line(0,-1){1}}
   \put(-.7071,-.7071){\line(1,0){1}}
   \put(0,1){\line(-1,-1){.7071}}
   \put(1,0){\line(-1,-1){.7071}}
   \put(-.7071,-.7071){\circle*{.1}}
   \put(-.7071,.2929){\circle*{.2}}
   \multiput(0,0)(1,0){2}{\circle*{.2}}
   \multiput(0,1)(1,0){2}{\circle*{.2}}
   \multiput(-.7071,-.7071)(1,0){2}{\circle*{.2}}
   \multiput(0,0)(.25,0){4}{\circle*{.03}}
   \multiput(0,0)(0,.25){4}{\circle*{.03}}
   \multiput(0,0)(-.1768,-.1768){4}{\circle*{.03}}
   \end{picture}}}
\newcommand{\twoplaq}{\setlength{\unitlength}{1cm}\raisebox{-.5cm}{
   \begin{picture}(1.2,1.2)(-.6,-.6)
   \basispl
   \put(-.5,-.5){\circle*{.1}}
   \put(-.5,.5){\circle*{.1}}
   \put(.5,-.5){\circle*{.1}}
   \put(.5,.5){\circle*{.1}}
   \put(0,-.5){\circle*{.1}}
   \put(0,.5){\circle*{.1}}
   \put(.5,0){\circle*{.1}}
   \put(-.5,0){\circle*{.1}}
   \end{picture}}}

\begin{document}
\vskip-1cm
\hfill INLO-PUB-06/96
\vskip4mm

\title{Testing Improved Actions\ \footnote{Talk presented by the last author
at the second workshop ``Continuous advances in QCD'', Univ. of Minnesota,
March 28-31, 1996.}}

\author{Margarita Garc\'{\i}a P\'erez, Jeroen Snippe and Pierre van Baal}

\address{Intsituut-Lorentz for Theoretical Physics, University of Leiden,
P.O.Box 9506, NL-2300 RA Leiden, The Netherlands}

\maketitle\abstracts{We discuss testing improved actions in the
context of finite volume gauge theories, where both results for
the continuum and the Wilson lattice action are known analytically 
for volumes up to 0.7 fermi across. A new improved action is 
introduced, obtained by adding a $2\times2$ plaquette to the 
L\"uscher-Weisz Symanzik action, for which the gauge field propagator 
greatly simplifies. We call this the square Symanzik action. We present 
the tree-level parameters of this improved action and the value of its 
Lambda parameter. We also give some Monte Carlo results and discuss some 
of the issues related to violations of unitarity at the scale of the 
lattice cutoff due to next-to-nearest coupling in the time direction.}

\section{Introduction}
Lattice gauge theory~\cite{wil} has proven to be a viable tool for 
non-perturbative studies in field theories and non-Abelian gauge
theories in particular. In perturbation theory it can be 
proven~\cite{reisz,lue} to have a continuum limit, and all results 
indicate that non-perturbatively the same will hold~\cite{creu,m-m}. 
Despite the tremendous increase of computer power since the first 
Monte Carlo calculations~\cite{creu2}, it remains a technical
challenge to make the lattice spacing $a$ small and the physical volume 
large enough. In particular problems of critical slowing down at small 
values of the coupling, implying that the algorithm is not 
efficient in probing the relevant portions of field space, makes a 
straightforward approach of decreasing the lattice spacing (while 
keeping the physical volume large enough) a costly procedure. Instead 
one can use the increased computer power of today to study alternative 
lattice actions, which are chosen to remove as much as possible the 
scaling violations introduced by a finite lattice cutoff. The 
computational overhead in using a more complicated lattice action is 
usually not more than a few to ten times the cost for the 
Wilson action. 
\subsection{Symanzik improvement}
The notion of improved actions was introduced more than a decade ago and 
in particular the study of Symanzik~\cite{sym} for scalar theories and 
L\"uscher and Weisz~\cite{luwe} for non-Abelian gauge theories have been 
influential. The last few years have seen a surge 
in applying these ideas in actual simulations~\cite{lat95}. 
When considering a lattice action, here restricting ourselves to pure gauge 
theories, the dynamical variables live on the (directed) links of the lattice 
and are elements of the gauge group $U_\mu(x)$. The connection to a continuum 
configuration is provided in terms of parallel transport of the vector 
potential $A_\mu(x)$ along the link $x\ra x+\hat\mu$ ($\hat\mu$ the 
directional vectors of the hypercubic lattice)
\be
U_\mu(x)=P\exp(\int_0^a A_\mu(x+s\hat\mu)ds)\ .
\ee
This allows us to express any of the lattice actions in powers of the
lattice spacing. Terms that are of order $a^4$ (in four dimensions), 
being the volume of a unit cell on the lattice, correspond to the naive 
continuum limit. Higher order terms of which we will quote the powers
of $a$ relative to this volume factor, correspond to irrelevant operators.
For example, for the Wilson action one finds the result~\cite{mgp}
\bea
S&=&\sum_{P}\Tr(1-U(P))=\sum_{x,\mu,\nu}\Tr\left(1-\plaq\right)\nonumber\\
&=&\sum_{x,\mu,\nu}\Tr(1-U_\mu(x)U_\nu(x+\hat{\mu})U_\mu^{\dagger}(x+\hat{\nu})
U_\nu^{\dagger}(x))\nonumber\\
&=&\sum_{x,\mu,\nu}-\Tr[\frac{a^4}{2}F_{\mu\nu}^2(x)-\frac{a^6}{24}\left(
(\cD_\mu F_{\mu\nu}(x))^2+(\cD_\nu F_{\mu\nu}(x))^2\right)+\frac{a^8}{24}
\{F_{\mu\nu}^4(x)\nonumber\\& &\hskip1.5cm+\frac{1}{30}\left((\cD_\mu^2 
F_{\mu\nu}(x))^2+(\cD_\nu^2 F_{\mu\nu}(x))^2\right)+\frac{1}{3}\cD_\mu^2 
F_{\mu\nu}(x)\cD_\nu^2 F_{\mu\nu}(x)\nonumber\\& &\hskip1.5cm
-\frac{1}{4}(\cD_\mu \cD_\nu F_{\mu\nu}(x))^2\}]+\cO (a^{10})
\quad,\label{eq:wac}
\eea
Note that apart from the corrections to the action density, sums need
to be converted to integrals (which for smooth fields would only give
exponential corrections, but for rough fields is part of the 
renormalization procedure) and the measure for integration over the
fields needs to be converted from (compact) link variables to vector 
potentials. The main advantage of the lattice formulation is of course 
its intrinsic gauge invariance. In the continuum, even in a finite volume, 
the integral over the irrelevant gauge degrees of variables is hard to 
define unambiguously. We note that modifications in the integration measure 
can, in principle, be absorbed in the action. 

To compensate for the irrelevant higher dimensional operators in the action 
one can follow two routes. One is using Wilson's renormalization group,
introducing blocking transformations which for finite couplings and lattice 
spacings define a renormalized trajectory along which the physical quantities 
remain constant~\cite{reng} (giving a perfect lattice action). Its limit for 
zero coupling is the classically perfect lattice action~\cite{has}. In a 
sense this can be seen, at least in principle, as a tree-level improved action 
to all orders in the lattice spacing, but through the blocking transformations
is supposed to also correct for the other errors we mentioned above. It may be 
expected, as demonstrated to one-loop order in the O(3) model~\cite{farch}, that
power-like cut-off effects in loop corrections are absent too. In this should 
lie the strength of classically perfect lattice actions, despite the need for 
truncations in actual numerical calculations. A large redundancy in the way 
one parametrizes these actions is reflected in the freedom of choosing the 
blocking transformations. One will try to keep the action as local as possible.

The large redundancy is also clear in the Symanzik improvement scheme,
which is based on perturbative computations of physical quantities.
At tree-level there are many ways of choosing the action in terms
of the gauge invariant Wilson loops that extend further than a
single plaquette, such that the order $a^2$ correction has a vanishing 
coefficient. The class of actions we will consider is defined by 
($<>$ implies averaging over orientations~\cite{mgp})
\bea
S(\{c_i\})&\equiv&\sum_x\Tr\{c_0\left\langle\sn\plaqa\,\right\rangle
+2c_1\left\langle\sn\twooneplaq\,\right\rangle
+\frac{4}{3}c_2\left\langle\sn\cornplaq \,\right\rangle
\nonumber\\&&
+\ 4c_3\left\langle\sn\hookplaq\,\right\rangle
+c_4\left\langle\sn\twoplaq \,\right\rangle\}\nonumber\\
&=&-\frac{a^4}{2}(c_0+8c_1+8c_2+16c_3+16c_4)\sum_{x,\mu,\nu}\Tr(F_{\mu\nu}^2
(x))\nonumber\\&&+\frac{a^6}{12}(c_0+20c_1-4c_2+4c_3+64c_4)\sum_{x,\mu,\nu}
\Tr(\cD_{\mu}F_{\mu\nu}(x))^2\nonumber\\
&&+a^6(\frac{c_2}{3}+c_3)\sum_{x,\mu,\nu,\lambda}
\Tr(\cD_{\mu}F_{\mu\lambda}(x)\cD_{\nu}F_{\nu\lambda}(x))\nonumber\\
&&+a^6\frac{c_2}{3}\sum_{x,\mu,\nu,\lambda}
\Tr((\cD_\mu F_{\nu\lambda})^2)+\cO(a^8)\ .
\label{eq:weisac}
\eea
We have added a $2\times2$ plaquette to the parametrization of the 
lattice action employed in ref.~8. (Note that sometimes
in the literature conventions are used where $c_2$ and $c_3$ are 
interchanged, e.g. refs.~10,14). We will motivate further on 
the advantage of adding this new plaquette. Using this action in 
perturbative calculations, one finds that the coefficients $c_i$ need 
to be corrected at higher order in $g_0^2$. This is achieved by computing 
a physical quantity to the required order and impose the vanishing of 
the $\cO(a^2)$ term to the relevant order in the coupling. This is 
called on-shell improvement and should be independent of the set of 
quantities required to remove the $\cO(a^2)$ terms. Note that some 
redundancies appear due the invariance under field redefinitions. As 
was shown by L\"uscher and Weisz~\cite{onsh} this allows one to choose 
$c_3\equiv 0$.

\subsection{Tadpole improvement}
The most urgent question would be how small the coupling constant (and 
hence the lattice spacing) has to be chosen to be able to truncate the 
perturbative construction of these improved actions. Relevant for 
questions related to truncating a perturbative series is which definition 
of the coupling constant to take. Formally, lattice perturbation theory is 
defined in the bare coupling constant. Usually, however, it helps greatly 
to convert the coupling constant to one that is defined at the scale of 
the process for which one is computing the perturbative corrections. This 
works well in the computation of physical quantities. However, one may 
doubt if this is valid for the computations of the improved action, as 
the corrections are by definition not physical. Still, a coupling at the 
scale of the cutoff would presumably perform better than the bare coupling. 
It has become customary to choose for this Parisi's mean field 
coupling~\cite{par}, extracted from the plaquette expectation value,
\be
g_P^2\equiv-a(c_i,N)\log(u_0),\quad u_0=<{1\over N}{\rm Re}\Tr U(P)>^{1/4}
\ee
where $a$ is a constant, defined such that $g_P=g_0+\cO(g_0^3)$. It is clear 
that this is just one possible choice. Nevertheless in terms of this coupling, 
deviations from asymptotic scaling seem to be much smaller than in 
terms of the bare lattice coupling. One usually relates this coupling 
to one that is obtained by a resummation of tadpole diagrams, although 
this is hard to make precise~\cite{periw}. Similarly one may attempt to 
apply ``mean field'' type corrections to the improvement coefficients 
that take into account the apparent large renormalization factors for 
the links, implied by the large difference between $g_P$ and $g_0$ at 
moderate and large couplings. This has been known as tadpole improvement 
and is observed to work very well, in particular when considering the 
rotational invariance of the heavy quark potential~\cite{lema}. 
It should be emphasized that tadpole improvement can be seen as a 
rearrangement of perturbation theory. When only tree-level improvement 
coefficients are known the coefficients $c_i$ are divided 
by a factor $u_0$ for each extra link (compared to the single 
plaquette).
\be
c_1\ra c_1/u_0^2,\ c_2\ra c_2/u_0^2,\ c_3\ra c_3/u_0^2,\ c_4\ra c_4/u_0^4
\ee
For the only case so far where the one-loop improvement corrections to 
$c_i$ are known, one adjusts~\cite{corn} the coefficients according to 
the expansion of $g_P$ in the bare coupling $g_0$. The implicit assumption 
of tadpole improvement is that corrections of $\cO(g_P^4)$ can be neglected. 
Although in many cases tadpole improvement is shown to reduce the size of 
the perturbative corrections and to improve numerical results for very coarse 
lattices indeed~\cite{corn}, it remains a rather ad hoc procedure. As long as 
one needs to show the validity by comparing with the carefully extrapolated 
Wilson action results, one should shed doubt on the usefulness of this 
procedure in particular when one starts to ``fudge'' with alternate 
definitions of $u_0$ (for example extracted from more extended Wilson loops). 
What would be required is a well motivated definition of (a class of) improved 
lattice actions where similar to the Wilson action careful scaling studies 
are performed, that allow one to extrapolate results independently to the 
continuum and infinite volume. Working only at the coarsest lattices 
will probably prove to be insufficient to obtain reliable results.
In particular systematic studies that use finite-size techniques as developed 
by the alpha collaboration~\cite{alpha} for determining the running coupling 
constant and renormalization factors~\cite{nonp} seem mandatory for a careful 
study of the systematic errors involved.

\subsection{The setting}
Motivated by all these problems we set out to test improved actions in the 
context of finite volume spectroscopy. The clear advantage is that for volumes
below 0.7 fm, results for the low-lying spectrum are know more or less 
analytically~\cite{vb}, both for the continuum and for the Wilson action. 
In such a case we can be precise in saying how much improvement is achieved.
We can not reach the coarsest lattices employed so far~\cite{corn}. Our volume 
has to be at least three lattice spacings across, which means that below
0.7 fm we can only reliably reach lattice spacings of the order of 0.2 fm. 
The data presented here will be for much smaller lattice spacing, where 
nevertheless scaling violations in mass ratios are particularly big. 

In order to also study the results of the improved actions analytically
we introduced a new improved lattice action for which the gauge field
propagator simplifies greatly. The explicit form of the propagator also
will make clear that there are violations of unitarity (poles in the 
propagator with negative residue) at the scale of the cutoff. In the 
continuum limit these are harmless, in the same way as Pauli-Villars 
regulator fields are harmless below the scale of the cutoff. It does,
however, cause difficulties in extracting a Hamiltonian~\cite{ham}.

\section{The new improved action}
To study perturbation theory on the lattice we write $U_\mu(x)=\exp(q_\mu(x))$
and expand the lattice action to second order in $q$. At tree level we have,
as for the L\"uscher-Weisz Symanzik action, $c_2\!=\!c_3\!=\!0$, cf. Eq.~3. To 
remove the remaining $\cO(a^2)$ irrelevant operator we have the three constants
$c_0$, $c_1$ and $c_4$ at our disposal. One is eliminated by requiring the 
appropriate continuum limit without the need of rescaling the coupling constant.
For the LW Symanzik action, corresponding to $c_4\!=\!0$, one 
finds $c_0\!=\!5/3$ and $c_1\!=\!-1/12$. The disadvantage of this choice in 
perturbation theory becomes clear when writing the quadratic part of the action
\bea
S_2&=&\!\!\!\sum_{x,\mu,\nu}\!\!-\half\Tr[c_0(\partial_\mu q_\nu(x)\!-\!
\partial_\nu 
q_\mu(x))^2+2c_1\{(2+\partial_\mu)(\partial_\mu q_\nu(x)\!-\!\partial_\nu q_\mu(
x))\}^2\nonumber\\&&\hskip1cm+c_4\{(2+\partial_\nu)(2+\partial_\mu)(
\partial_\mu q_\nu(x)-\partial_\nu q_\mu(x))\}^2]\ ,
\eea
where the lattice derivative $\partial_\mu$ is the difference operator
\be
\partial_\mu\varphi(x)\equiv \varphi(x+\hat\mu)-\varphi(x)\ .
\ee
There is no ``covariant'' gauge condition that will make the gauge
field propagator diagonal in the space-time indices. In analytic
calculations this makes perturbative computations cumbersome, in
particular in the presence of a background field, but of course
not impossible~\cite{luwe2}. Nevertheless, if we choose 
\be
c_4\cdot c_0=c_1^2\ ,
\ee
a condition {\em invariant under} the tadpole modification in Eq.~5,
we can rearrange Eq.~6 by completing squares, from which we read off 
a simple gauge condition
\bea
S_2=&&\hskip-9mm-\!\sum_{x,\mu,\nu}\!\!c_0\Tr[\partial_\mu q_\nu(x)
\left(1\!+\!z(2\!+\!\partial_\mu^\dagger)(2\!+\!\partial_\mu)\right)
\left(1\!+\!z(2\!+\!\partial_\nu^\dagger)(2\!+\!\partial_\nu)\right)
\partial_\mu q_\nu(x)]\nonumber\\&+&\hskip-3mm\sum_x\Tr\cF_{gf}^2(x),\quad
\cF_{gf}(x)\equiv\sqrt{c_0}\sum_{\mu}\partial_\mu^\dagger
\left(1\!+\!z(2\!+\!\partial_\mu^\dagger)(2\!+\!\partial_\mu)\right)\!q_\mu(x),
\eea
where $z\equiv c_1/c_0$. At tree level we find $c_0=16/9$, $c_1=-1/9$, 
$c_2=c_3=0$ and $c_4=1/144$, whereas for $u_0\neq 1$ we have 
$z=-1/16u_0^2$ and $c_0=1/(1+4z)^2$. We propose to call this new action 
the {\em square} Symanzik action, or square action for short. We find the 
following propagators
\bea
{\rm Ghost}:&&P(k)=\frac{1}{\sqrt{c_0}\sum_\lambda\left(4\sin^2(k_\lambda/2)
+4z\sin^2k_\lambda\right)}\ ,\nonumber\\{\rm Vector}:&&P_{\mu\nu}(k)=
\frac{P(k)\delta_{\mu\nu}}{\sqrt{c_0}\left(1+4z\cos^2(k_\mu/2)\right)}\ .
\eea
This simple form of the propagator allows us to demonstrate a general
feature of improved actions, which have couplings in the time directions
that are not nearest neighbor. We introduce the standard lattice propagators
\bea
P_s&\equiv&\frac{1}{4\sin^2(k_0/2)+\omega_s^2}\ ,\ \omega_\pm^2\equiv
-\frac{(1+4z)}{2z}\left(1\pm\sqrt{1+\frac{4z\omega^2}{(1+4z)^2}}\right),
\nonumber\\\omega_0^2&\equiv&-\frac{1+4z}{z}\ ,\ \omega^2\equiv
4\sum_{i=1}^3\sin^2(k_i/2)\left(1+4z\cos^2(k_i/2)\right),
\eea
which have a single pole in the $k_0$ Brillouin zone with the standard 
residue as would have been obtained from the Wilson action (except that 
$\omega\equiv\omega(z\!=\!0)$ is replaced by $\omega_s$). This allows us 
to write
\bea
P&=&Z(P_--P_+),\ P_{jj}=\frac{(1+4z)}{\left(1+4z\cos^2(k_j/2)\right)}P,
\nonumber\\ P_{00}&=&(1+4z)\left(\frac{\omega_-^2Z}{\omega^2}P_-
-\frac{\omega_+^2Z}{\omega^2}P_++\frac{\omega_0^2}{\omega^2}P_0\right),
\eea
where the $Z$ factor is given by $Z\equiv1/\sqrt{1+4z\omega^2/(1+4z)^2}$,
showing that there are unphysical propagating negative (and positive) norm 
states. This is a direct consequence of the fact that the transfer matrix, 
as defined in ref.~23, is not hermitian. The unphysical
masses involved are at the scale of the cutoff (in lattice units $\cO(1)$). 
For $\vec k=\vec 0$ and $u_0=1$ one finds $\omega_-=0$ and $\omega_+=\omega_0
=\sqrt{12}$. Although the unphysical states are interacting in a complicated 
way, and do not just appear in closed loops (thereby making it somewhat 
misleading to call them ghosts) they can be shown not to lead to imaginary
eigenvalues and violations of unitarity in the low-lying spectrum~\cite{ham}. 
In particular the low-lying spectrum for gauge theories in an intermediate 
volume is obtained by integrating out the non-zero momentum modes in a loop 
expansion and solving the resulting effective theory using a Rayleigh-Ritz 
analysis which includes imposing proper boundary conditions in field space 
implied by the gauge invariance ~\cite{vb}. The effect of the spurious poles 
is integrated out at the same level of accuracy as the high momentum modes. 
Here we will only use our square Symanzik action to calculate the effective 
potential for an abelian constant background field, which greatly simplifies 
due to the diagonal structure of the gauge field propagator. Also it will
be used to compute the Lambda parameter.
\section{The effective potential}
The effective potential for a constant abelian background field $C_\mu$ is 
particularly simple to calculate since the background field will give rise 
to a shifted momentum, $k_\mu\ra k_\mu+sC_\mu/N_\mu$, where $N_\mu$ is the 
size of the lattice in each of the four directions. For the modes $q$ that 
commute with the background field one has $s=0$, whereas for the two charged 
components with respect to the abelian direction defined by the background 
field one takes $s=\pm1$. To one-loop order the effective potential is simply 
obtained by summing over the logarithm of the eigenvalues of the quadratic 
fluctuation operator, properly corrected for by the ghost contributions.
We take $C_0=0$ and $N_i=N$, such that
\be
V_1(C)=\frac{1}{N_0}\sum_{k,s}\{\half\sum_\mu\log\lambda_\mu(k+sC/N)-
\log\lambda_{gh}(k+sC/N)\}
\ee
The eigenvalues $\lambda(k)$ are directly read off from the explicit
expressions for the propagators in Eq.~10.
\bea
\lambda_{gh}(k)&=&\sqrt{c_0}\sum_\nu 4\sin^2(k_\nu/2)(1+4z\cos^2(k_\nu/2)),
\nonumber\\\lambda_\mu(k)&=&\sqrt{c_0}(1+4z\cos^2(k_\mu/2))\lambda_{gh}(k).
\eea
This formula holds equally well for the Wilson action (where $z\equiv0$). As 
follows from the factorization of the propagator we can write $\lambda_{gh}$
as the product of two eigenvalues. They will come in complex conjugate pairs 
when the momentum gets too close to the edge of the Brillouin zone. 
Nevertheless splitting $\log\lambda_{gh}$ in these separate contributions, that
depend on $k_0$ as in the background field calculation for the Wilson action, 
allows one to perform the sum over $k_0$ exactly~\cite{vb}. In particular
for $N_0\ra\infty$ one finds the following explicit result
\be
V_1(\vec C)=N\!\!\!\!\sum_{\vec n\in\zahlen^3}\!\Bigl\{
4{\rm asinh}\left(2u_0\sqrt{1\!+\!4z\!+\!\frac{\omega^2}{2}\!+
\!\omega\sqrt{1\!+\!\frac{\omega^2}{4}}}\right)+
\sum_i\log\left(\Omega_i\right)\Bigr\},
\ee
up to an irrelevant overall constant. Here $\Omega_i\equiv 1\!+\!4z\cos^2[
(2\pi n_i\!+\!C_i)/(2N)]$ and $\omega\equiv\omega(\vec k\!=\!(\pi\vec n\!+
\!\half\vec C)/N)$ as defined in Eq.~11. For $z\ra 0$ this gives the well known 
result obtained for the Wilson action, whereas $N\ra\infty$ recovers the 
result for the continuum theory~\cite{vb}. 
\begin{figure}{\tt}
\vspace{2in}
\includegraphics{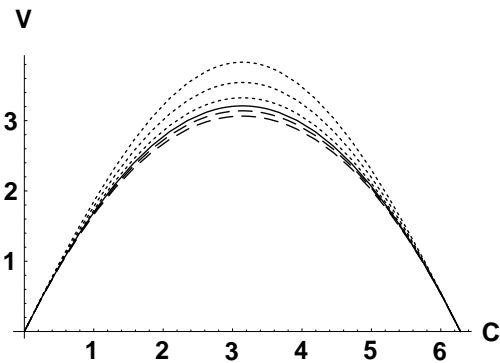}
\caption{The effective potential for a constant Abelian background
field $A_1=\half iC\sigma_3/N$. The full line represents the continuum 
result. The lower two dashed curves are for the square Symanzik action with 
$N=3$ and 4 ($N=6$ is indistinguishable from the continuum curve). The upper 
three dotted curves are for the Wilson action with $N=3,4$ and 6.}
\label{fig:veff}
\end{figure}
In figure 1 we compare the results of $V_1(C)$ for the square Symanzik action 
(lower two curves for $N=3$ and $N=4$), with those for the Wilson action 
(upper three curves for $N=3,4$ and 6). The full curve gives the result of 
$V_1(C)$ for the continuum (we have chosen $\vec C=(C,0,0)$). For the new 
improved action the result for $N=6$ can already not be distinguished from 
that in the continuum at the scale of this figure. One might even fear that 
choosing $u_0\neq1$ will make the agreement worse. However, the effective 
potential is not a spectral quantity and deviations of $V_1(C)$ from the 
continuum can in principle be compensated by other corrections in the effective 
Hamiltonian for the zero-momentum modes. 
\section{The Lambda parameter}
One can now follow the same strategy as in calculating the zero-momentum
effective Hamiltonian for the Wilson action. The difficulty lies in 
converting a non-hermitian transfer matrix, defined by the euclidean
path integral, to a Hamiltonian. We postpone this to a future publication, 
where it will be shown that by a suitable field redefinition one can map the
effective action with next-to-nearest neighbor couplings in the time 
direction to one with nearest neighbor coupling. After this transformation
(whose Jacobian will give rise to corrections of odd powers in the lattice 
spacing), one can convert the path integral through a hermitian transfer
matrix to a Hamiltonian~\cite{vb} (this conversion will introduce 
corrections that are of even powers in the lattice spacing). At first, finding 
scaling violations that are of odd powers in the lattice spacing
may seem rather puzzling, but it turns out they are required to {\em exactly 
cancel} similar scaling violations that appear in some coefficients of 
the effective action~\cite{lat96}. 

However, to determine the Lambda parameter for the square Symanzik action,
it is sufficient to determine the effective action and study the finite 
difference in the renormalization as compared to (say) the Wilson action 
in the limit of zero lattice spacing. Although in the renormalized action 
the tree level kinetic and potential terms, $\half(\partial_t c^a_i)^2$ and 
$\half(F_{ij}^a)^2$, will have coefficients that differ by finite 
renormalizations (due to the breaking of Lorentz invariance), the difference 
of these finite renormalizations between the different regularizations is 
expected to be the same for both terms. It should be noted that in principle 
one can perform a rescaling of the fields, but this would upset the 
periodicity (a consequence of gauge invariance) of the effective theory 
along the abelian constant potential. Indeed in background field 
calculations no field renormalizations appear.

Consequently two numerically independent determinations of the Lambda
parameter can be extracted from this background field calculation, as
for the Wilson action~\cite{vb}. In addition one can follow the 
alternative determination of the Lambda parameter through the heavy 
quark potential~\cite{wei}. All three determinations gave to a high
degree of accuracy the same result for SU(2). The result for SU(3)
was obtained with the latter method only. 
\bea
\Lambda_{S^2}/\Lambda_W&=&4.0919901(1)\quad {\rm for\ SU(2)},\nonumber\\
\Lambda_{S^2}/\Lambda_W&=&5.2089503(1)\quad {\rm for\ SU(3)}.
\eea
The index ${S^2}$ stands for the square Symanzik and $W$ for the Wilson 
action. Details of both methods to compute these 
ratios will be presented elsewhere. 
\section{Monte Carlo data}
In figure 2 we present the SU(2) Monte Carlo results for mass ratios in a 
small volume. Full details of the analysis will be presented elsewhere.
We chose the volume such that the lattice artefacts in the mass ratios,
the difference between the full curves (continuum) and the dotted curves 
(Wilson action for a $4^3\times\infty$ lattice), were maximal~\cite{vb}. 
For the Wilson action, at $\beta=4/g_0^2=3$ this corresponds
to a lattice spacing of approximately 0.018 fm. Another reason to pick
these parameters was to compare with earlier high precision data for the
Wilson action by Michael~\cite{mic} (crosses in the figure), so as to 
make sure no errors were made in the implementation and in the measurements 
of the masses. We also improved here somewhat on the statistical errors.
The data corresponding to the LW Symanzik action is represented by the triangles
and for our new square Symanzik action by the squares. 
In both cases we used tree-level
improvement only. The improvement is significant. For the LW Symanzik action 
the data is within two sigma of the continuum values. The results seem to 
indicate that the square Symanzik action is somewhat less effective, although
the difference is not significant. A comparison at coarser lattices will be
more interesting. The lattice results for the new action can be shown to 
confirm the value of the Lambda parameter computed from perturbation theory.
The value of $\beta$ for the Monte Carlo analysis with the square 
Symanzik action 
was chosen before the the perturbative calculations were completed.
\begin{figure}{\tt}
\vspace{3.75in}
\includegraphics{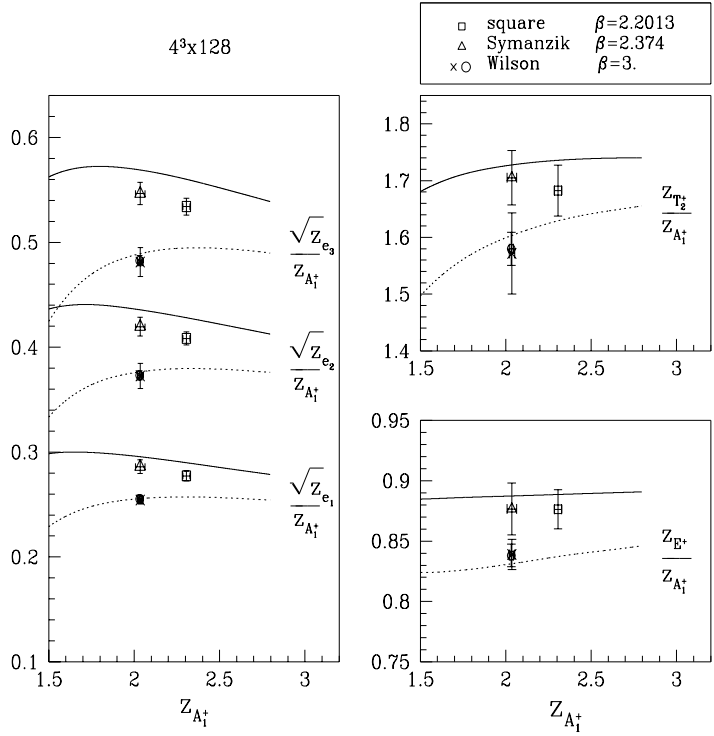}
\caption{SU(2) Monte Carlo data for mass ratios in a small volume on a 
lattice of size $4^3\times128$, using the Wilson action (crosses 
from existing data of Michael), the LW Symanzik action and our new square
Symanzik action. The lines give the analytic results: full for the continuum 
and dotted for the standard lattice action ($N=4$).}
\label{fig:MCdata}
\end{figure}

The mass ratios shown are all with respect to the scalar mass in the 
singlet representation $A_1^+$ of the cubic group. On the left we plot
the square root of the finite volume ``string tension'' for one, two and
three units of electric flux, whereas on the right we consider the tensor
doublet $E^+$ and triplet $T_2^+$, which become degenerate in large volumes
when rotational invariance is restored. The finite volume ``string tension''
is simply defined by dividing the energy of the electric flux states (also 
called torelon masses~\cite{mic}) by the length of the box. The electric flux
is defined with the help of twisted boundary conditions~\cite{tho}. The 
Wilson loop operator that closes due to the periodic boundary conditions can
be seen to create electric flux. Two (or three) units of electric flux
are obtained from loops that wind around two (or three) directions of the torus.
\section{Outlook}
Our contribution has been modest as we have not probed very coarse lattices. 
The Monte Carlo data presented here were for a lattice spacing of approximately 
0.02 fm. New Monte Carlo data will be presented elsewhere, studying lattice 
spacings of approximately 0.12 fm. We would also like to stress that we 
used mass ratios to test improvement. It has been well know that the Parisi 
mean field coupling constant improves the approach to asymptotic scaling 
quite well. We see this as a separate issue, quite (but not completely) 
unrelated to the issue of scaling. The reason is that the scale in lattice 
calculations is usually set by one of the masses (in pure gauge theories by
the string tension) anyhow.

From the practical point of view we presented an alternative action,
motivated by the requirement to simplify the perturbative calculations,
rather than to minimize the lattice artefacts. Nevertheless, comparisons
will give a clue on how big the higher order corrections are. Of course
at tree level it is trivial to write down many types of improved actions,
so we have set out to compute the one-loop improvement coefficients too,
to bring our new square Symanzik improved action to the same footing as 
the L\"uscher-Weisz Symanzik improved action. These results will be 
presented elsewhere. This allows one to test the ``universality'' of tadpole 
improvement. Again we wish to stress that the most interesting tests are 
those for the scaling of mass ratios.

\section*{Acknowledgments}
One of the authors (PvB) is grateful to the organizers of this workshop
for their generous invitation. He also thanks Mark Alford, Andreas Kronfeld, 
Peter Lepage, Paul Mackenzie, Uwe-Jens Wiese and Peter Weisz for fruitful 
discussions and correspondence at various occasions. MGP thanks Mike Teper 
for a useful correspondence. This work was supported in part by a grant from 
FOM and from NCF for use of the national Cray facility at SARA.

\end{document}